\documentclass[aps,pre,reprint,superscriptaddress,showpacs,preprintnumbers,amsmath,amssymb]{revtex4-1}
\usepackage{graphicx}
\graphicspath{{fig/}}
\usepackage{dcolumn}
\usepackage{bm}
\usepackage{color}
\begin{document}
\preprint{Preprint}
\title{Critical scaling of diffusion coefficients and size of rigid clusters of soft athermal particles under shear}
\author{Kuniyasu Saitoh}
\affiliation{Research Alliance Center for Mathematical Sciences, Tohoku University, 2-1-1 Katahira, Aoba-ku, Sendai 980-8577, Japan}
%
\author{Takeshi Kawasaki}
\affiliation{Department of Physics, Nagoya University, Nagoya 464-8602, Japan}
\date{\today}
\begin{abstract}
We numerically investigate the self-diffusion coefficient and correlation length of the rigid clusters (i.e., the typical size of the collective motions) in sheared soft athermal particles.
Here we find that the rheological flow curves on the self-diffusion coefficient are collapsed by the proximity to the jamming transition density.
This feature is in common with the well-established critical scaling of flow curves on shear stress or viscosity.
We furthermore reveal that the divergence of the correlation length governs the critical behavior of the diffusion coefficient,
where the diffusion coefficient is proportional to the correlation length and the strain rate for a wide range of the strain rate and packing fraction across the jamming transition density.
\end{abstract}
\maketitle
%
\section{Introduction}
Transport properties of soft athermal particles,\ e.g.\ emulsions, foams, colloidal suspensions, and granular materials, are important in science and engineering technology \cite{bird}.
In many manufacturing processes, these particles are forced to flow through pipes, containers, etc.
Therefore, the transportation of ``flowing particles" is of central importance for industrial applications \cite{larson}
and thus there is a need to understand how the transport properties are affected by the rheology of the particles.

Recently, the rheological flow behavior of soft athermal particles has been extensively studied by physicists
\cite{rheol0,pdf1,rheol1,rheol2,rheol3,rheol4,rheol5,rheol6,rheol7,rheol8,rheol9,rheol10,rheol11,rheol12,rheol13}.
It has been revealed that the rheology of such particulate systems depends not only on shear rate $\dot{\gamma}$ but also on packing fraction of the particles.
If the packing fraction $\phi$ is lower than the so-called \emph{jamming transition density} $\phi_J$ \cite{gn5},
steady state stress is described by either Newtonian \cite{rheol0,pdf1} or Bagnoldian rheology \cite{rheol1,rheol2,rheol3,rheol4,rheol5} (depending on whether particle inertia is significant).
However, if the packing fraction exceeds the jamming point, one observes yield stress at vanishing strain rate \cite{review-rheol0}.
These two trends are solely determined by the proximity to jamming $|\Delta\phi|\equiv|\phi-\phi_J|$ \cite{rheol0},
where rheological flow curves of many types of soft athermal particles have been explained by the critical scaling near the jamming transition
\cite{pdf1,rheol1,rheol2,rheol3,rheol4,rheol5,rheol6,rheol7,rheol8,rheol9,rheol10,rheol11}.

On the other hand, the mass transport or \emph{self-diffusion} of soft athermal particles seems to be controversial.
As is the rheological flow behavior on shear stress or viscosity, the diffusivity of the particles under shear is also dependent on both $\dot{\gamma}$ and $\phi$.
Its dependence on $\dot{\gamma}$ is weakened with the increase of $\dot{\gamma}$,\
i.e.\ the diffusivity $D$ exhibits a crossover from a linear scaling $D\sim\dot{\gamma}$ to the sub-linear scaling $D\sim\dot{\gamma}^q$ at a characteristic shear rate $\dot{\gamma}_c$,
where the exponent is smaller than unity, $q<1$ \cite{diff_shear_md7,diff_shear_md6,dh_md2,diff_shear_md2,diff_shear_md3,diff_shear_md4}.
For example, in molecular dynamics (MD) simulations of Durian's bubble model in two dimensions \cite{diff_shear_md7,diff_shear_md6}
and frictionless granular particles in three dimensions \cite{dh_md2},
the diffusivity varies from $D\sim\dot{\gamma}$ ($\dot{\gamma}<\dot{\gamma}_c$) to $D\sim\dot{\gamma}^{0.8}$ ($\dot{\gamma}>\dot{\gamma}_c$).
These results also imply that the diffusivity does not depend on spatial dimensions.
However, another crossover from $D\sim\dot{\gamma}$ to $D\sim\dot{\gamma}^{1/2}$ was suggested by the studies of amorphous solids
(though the scaling $D\sim\dot{\gamma}^{1/2}$ is the asymptotic behavior in rapid flows $\dot{\gamma}\gg\dot{\gamma}_c$) \cite{diff_shear_md2,diff_shear_md3,diff_shear_md4}.
In addition, it was found in MD simulations of soft athermal disks that,
in a sufficiently small flow rate range, the diffusivity changes from $D\sim\dot{\gamma}$ ($\phi<\phi_J$) to $\dot{\gamma}^{0.78}$ ($\phi\simeq\phi_J$) \cite{diff_shear_md0},
implying that the crossover shear rate $\dot{\gamma}_c$ vanishes as the system approaches jamming from below $\phi\rightarrow\phi_J$.

Note that the self-diffusion of soft athermal particles shows a clear difference from the diffusion in glass;
\emph{no plateau} is observed in (transverse) mean square displacements (MSDs) \cite{diff_shear_md0,diff_shear_md2,diff_shear_md3,diff_shear_md4,diff_shear_md7,dh_md2}.
The absence of sub-diffusion can also be seen in quasi-static simulations ($\dot{\gamma}\rightarrow 0$) of soft athermal disks \cite{dh_qs1}
and MD simulations of granular materials sheared under constant pressure \cite{diff_shear_md1}.

Because the self-diffusion can be associated with collective motions of soft athermal particles,
physicists have analyzed spatial correlations of velocity fluctuations \cite{rheol0} or non-affine displacements \cite{nafsc2} of the particles.
Characteristic sizes of collectively moving regions,\ i.e.\ \emph{rigid clusters}, are then extracted as a function of $\dot{\gamma}$ and $\phi$.
However, there is a lack of consensus on the scaling of the sizes.
For example, the size of rigid clusters $\xi$ diverges as the shear rate goes to zero $\dot{\gamma}\rightarrow 0$ so that the power-law scaling $\xi\sim\dot{\gamma}^{-s}$ was suggested,
where the exponent varies from $s=0.23$ to $0.5$ depending on numerical models and flow conditions \cite{dh_md2,diff_shear_md1}.
The dependence of $\xi$ on $\phi$ is also controversial.
If the system is below jamming, the critical scaling is given by $\xi\sim|\Delta\phi|^{-w}$,
where different exponents in the range between $0.5\le w\le 1.0$ have been reported by various simulations \cite{lng0,rheol0,nafsc2,rheol16}.
In contrast, if the system is above jamming, the size becomes insensitive to $\phi$ (or exceeds the system size $L$) as only $L$ is the relevant length scale,\
i.e.\ $\xi\sim L$, in a quasi-static regime \cite{diff_shear_md2,diff_shear_md3,diff_shear_md4,pdf1}.

From a scaling argument \cite{diff_shear_md1,dh_qs1}, a relation between the diffusivity and size of rigid clusters was proposed as
\begin{equation}
D\sim d_0\xi\dot{\gamma}~,
\label{eq:rigid_cluster}
\end{equation}
where $d_0$ is the particle diameter.
It seems that previous results above jamming,\
i.e.\ as $\dot{\gamma}$ increases, $D/\dot{\gamma}$ changes from constant to $\dot{\gamma}^{-1/2}$ and corresponding $\xi$ undergoes from $L$ to $\dot{\gamma}^{-1/2}$,
support this argument \cite{diff_shear_md2,diff_shear_md3,diff_shear_md4}.
However, the link between the diffusivity and rigid clusters \emph{below jamming} is still not clear.

In this paper, we study the self-diffusion of soft athermal particles and the size of rigid clusters.
The particles are driven by simple shear flows and their fluctuating motions around a mean velocity field are numerically calculated.
From numerical results, we extract the diffusivity of the particles and explain its dependence on the control parameters,\ i.e.\ $\dot{\gamma}$ and $\phi$.
We investigate wide ranges of the control parameters in order to unify our understanding of the diffusivity in both slow and fast flows
($\dot{\gamma}<\dot{\gamma}_c$ and $\dot{\gamma}>\dot{\gamma}_c$) and both below and above jamming ($\phi<\phi_J$ and $\phi>\phi_J$).
Our main result is the critical scaling of the diffusivity $D$, which parallels the critical scaling of the size of rigid clusters $\xi$.
We find that the linear relation between $D$ and $\xi$ [Eq.\ (\ref{eq:rigid_cluster})] holds over the whole ranges of $\dot{\gamma}$ and $\phi$ if finite-size effects are not important.
In the following, we show our numerical method in Sec.\ \ref{sec:method} and numerical results in Sec.\ \ref{sec:result}.
In Sec.\ \ref{sec:disc}, we discuss and conclude our results and future works.
%
\section{Methods}
\label{sec:method}
We perform MD simulations of two-dimensional disks.
In order to avoid crystallization of the system, we randomly distribute an equal number of small and large disks
(with diameters $d_S$ and $d_L=1.4d_S$) in a $L\times L$ square periodic box \cite{gn1}.
The total number of disks is $N=8192$ and the packing fraction of the disks $\phi$ is controlled around the jamming transition density $\phi_J\simeq0.8433$ \cite{rheol0}.
We introduce an elastic force between the disks, $i$ and $j$, in contact as $\bm{f}_{ij}^\mathrm{e}=k\delta_{ij}\bm{n}_{ij}$,
where $k$ is the stiffness and $\bm{n}_{ij}\equiv\bm{r}_{ij}/|\bm{r}_{ij}|$ with the relative position $\bm{r}_{ij}\equiv\bm{r}_i-\bm{r}_j$ is the normal unit vector.
The elastic force is linear in the overlap $\delta_{ij}\equiv R_i+R_j-|\bm{r}_{ij}|>0$, where $R_i$ ($R_j$) is the radius of the disk $i$ ($j$).
We also add a damping force to every disk as $\bm{f}_i^\mathrm{d}=-\eta\left\{\bm{v}_i-\bm{u}(\bm{r}_i)\right\}$,
where $\eta$, $\bm{v}_i$, and $\bm{u}(\bm{r})$ are the damping coefficient, velocity of the disk $i$, and external flow field, respectively.
The stiffness and damping coefficient determine a time scale as $t_0\equiv\eta/k$.

\begin{figure}
\includegraphics[width=\columnwidth]{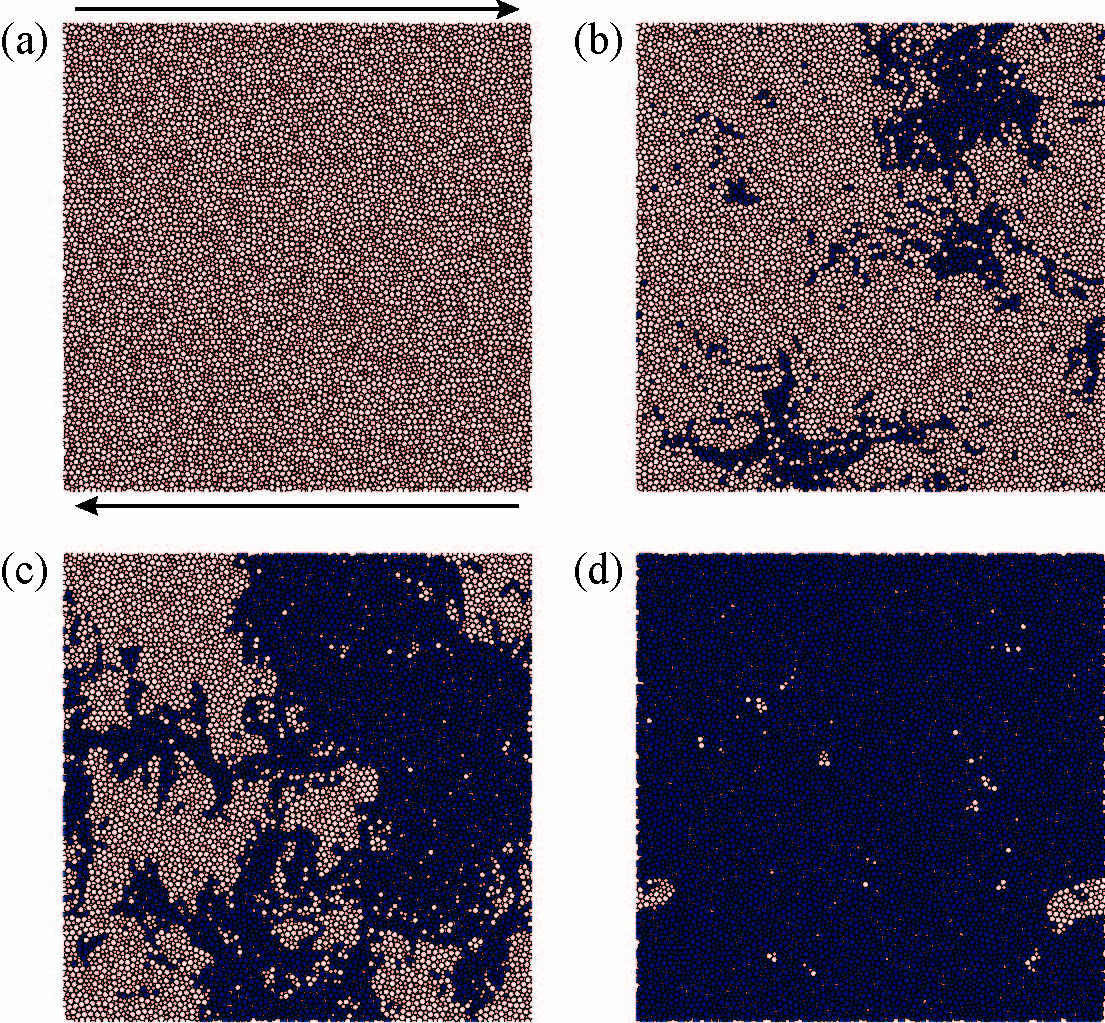}
\caption{
Snapshots of our MD simulations, where the particles (circles) are sheared as indicated by the horizontal arrows in (a).
The filled (open) circles represent mobile (immobile) particles, where the magnitude of their non-affine velocities $|\bm{v}_i-\bm{u}(\bm{r}_i)|$ is larger (smaller) than $10d_0\dot{\gamma}$.
The shear rate is $\dot{\gamma}=10^{-7}t_0^{-1}$ and the packing fraction $\phi$ increases as (a) $0.80$, (b) $0.84$, (c) $0.85$, and (d) $0.90$.
\label{fig:rigid}}
\end{figure}

To simulate simple shear flows of the system, we impose the external flow field $\bm{u}(\bm{r})=(\dot{\gamma}y,0)$ under the Lees-Edwards boundary condition \cite{lees},
where $\dot{\gamma}$ is the shear rate.
Motions of the disks are described by overdamped dynamics \cite{rheol0,rheol7,pdf1},\ i.e.\ $\sum_{j\neq i}\bm{f}_{ij}^\mathrm{e}+\bm{f}_i^\mathrm{d}=\bm{0}$,
so that we numerically integrate the disk velocity $\bm{v}_i=\bm{u}(\bm{r}_i)+\eta^{-1}\sum_{j\neq i}\bm{f}_{ij}^\mathrm{el}$ with a time increment $\Delta t = 0.1t_0$.
Figure \ref{fig:rigid} displays snapshots of our MD simulations, where the system is sheared along the horizontal arrows in (a).
Here, the shear rate is fixed to $\dot{\gamma}=10^{-7}t_0^{-1}$ and the packing fraction increases from $\phi=0.8$ (a) to $0.9$ (d).
We observe that, as $\phi$ increases, the magnitude of \emph{non-affine velocities}, $|\bm{v}_i-\bm{u}(\bm{r}_i)|$, is pronounced
and mobile particles (satisfying $|\bm{v}_i-\bm{u}(\bm{r}_i)|>10d_0\dot{\gamma}$ with the mean disk diameter $d_0\equiv(d_S+d_L)/2$) form clusters (filled circles).
The size of these clusters, or \emph{rigid clusters}, increases with $\phi$ [Figs.\ \ref{fig:rigid}(b) and (c)] and reaches the box size $L$ if the system is far above jamming [Fig.\ \ref{fig:rigid}(d)].

In the following, we analyze the data in a steady state, where shear strain applied to the system is larger than unity, and scale every time and length by $t_0$ and $d_0$, respectively.
%
%
\section{Results}
\label{sec:result}
In this section, we show our numerical results of the self-diffusion of soft athermal particles (Sec.\ \ref{sub:diff}).
We also extract rigid clusters (Fig.\ \ref{fig:rigid}) from numerical data in order to relate their sizes to the diffusivity (Sec.\ \ref{sub:rigid}).
We explain additional data of the rheology and non-affine displacements in Appendixes.
%
\subsection{Diffusion}
\label{sub:diff}
\begin{figure}
\includegraphics[width=\columnwidth]{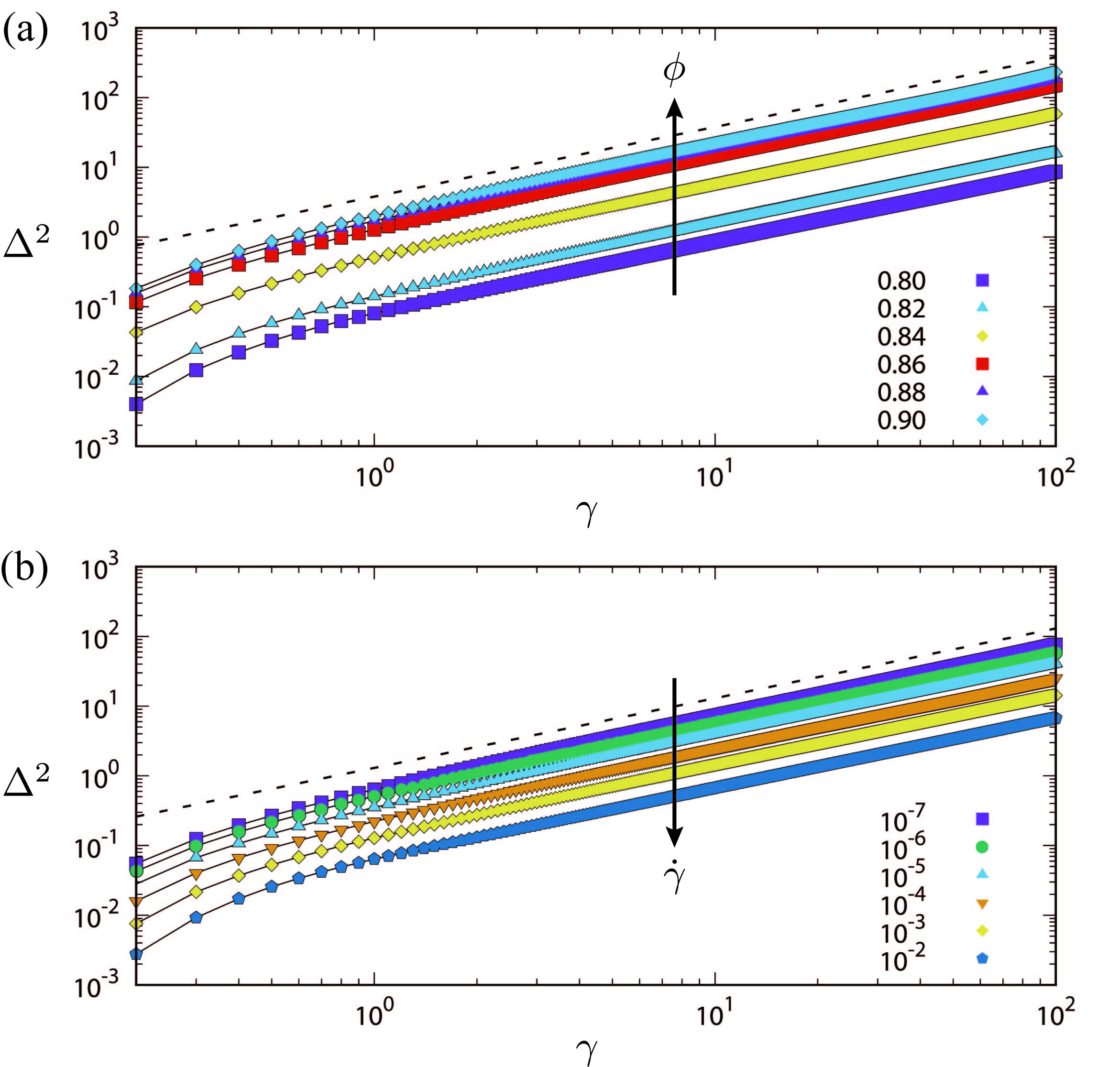}
\caption{
The transverse MSDs $\Delta^2$ [Eq.\ (\ref{eq:MSD})] as functions of the shear strain $\gamma\equiv\dot{\gamma}\tau$.
(a) The packing fraction $\phi$ increases as indicated by the arrow and listed in the legend, where the shear rate is $\dot{\gamma}=10^{-6}t_0^{-1}$.
(b) The shear rate $\dot{\gamma}$ increases as indicated by the arrow and listed in the legend, where the packing fraction is $\phi=0.84$.
\label{fig:msdy}}
\end{figure}

We analyze the self-diffusion of soft athermal particles by the transverse component of \emph{mean squared displacement} (MSD) \cite{diff_shear_md0,diff_shear_md1,diff_shear_md3,diff_shear_md4},
\begin{equation}
\Delta(\tau)^2 = \left\langle\frac{1}{N}\sum_{i=1}^N\Delta y_i(\tau)^2\right\rangle~.
\label{eq:MSD}
\end{equation}
Here, $\Delta y_i(\tau)$ is the $y$-component of particle displacement and the ensemble average $\langle\dots\rangle$ is taken over different choices of the initial time (see Appendix \ref{sec:nona} for the detail)
\footnote{The MSDs defined by the \emph{total} non-affine displacements show quantitatively the same results (data are not shown).}.
Figure \ref{fig:msdy} displays the MSDs [Eq.\ (\ref{eq:MSD})] with different values of (a) $\phi$ and (b) $\dot{\gamma}$.
The horizontal axes are the time interval scaled by the shear rate, $\gamma\equiv\dot{\gamma}\tau$,\ i.e.\ the shear strain applied to the system for the duration $\tau$.
As can be seen, every MSD exhibits a crossover to the normal diffusive behavior, $\Delta(\tau)^2\sim\dot{\gamma}\tau$ (dashed lines),
around a crossover strain $\gamma=\gamma_c\simeq 1$ regardless of $\phi$ and $\dot{\gamma}$.
The MSDs below jamming ($\phi<\phi_J$) monotonously increase with the increase of packing fraction,
while they (almost) stop increasing once the packing fraction exceeds the jamming point ($\phi>\phi_J$) [Fig.\ \ref{fig:msdy}(a)].
The dependence of MSDs on the shear rate is monotonous; their heights decrease with the increase of $\dot{\gamma}$ [Fig.\ \ref{fig:msdy}(b)].
These trends well correspond with the fact that the non-affine displacements are amplified in slow flows of dense systems,\ i.e.\ $\dot{\gamma}t_0\ll 1$ and $\phi>\phi_J$ \cite{saitoh11}.
In addition, different from thermal systems under shear \cite{rheol10,nafsc5,th-dh_md1}, any plateaus are not observed in the MSDs.
Therefore, neither ``caging" nor ``sub-diffusion" of the particles exists in our sheared athermal systems \cite{dh_md2,dh_qs1,dh_md1}.
%

\begin{figure}
\includegraphics[width=\columnwidth]{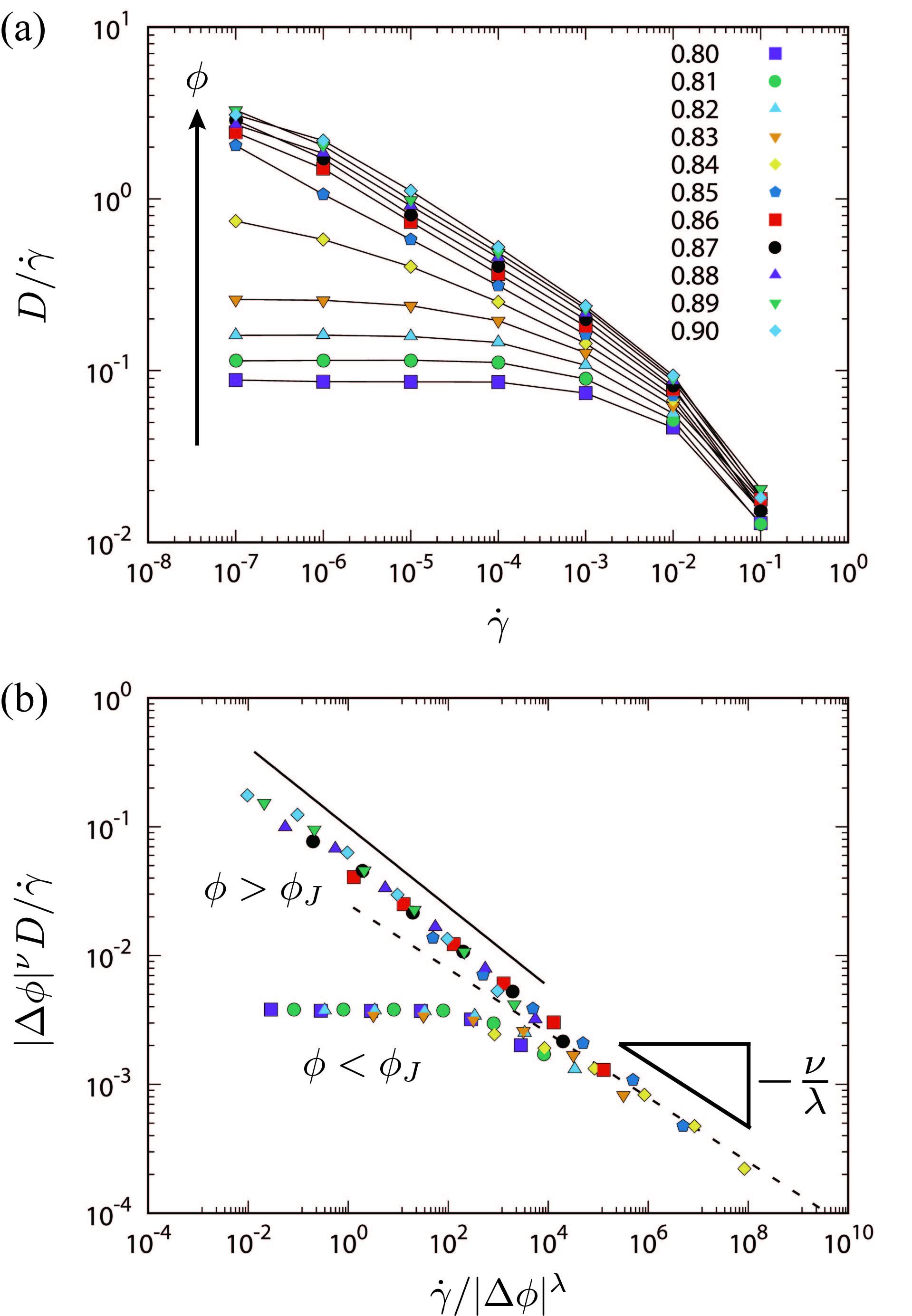}
\caption{
(a) The diffusivity over the shear rate, $D/\dot{\gamma}$, as a function of $\dot{\gamma}$, where $\phi$ increases as indicated by the arrow and listed in the legend.
(b) \emph{Scaling collapses} of the diffusivity, where $\Delta\phi\equiv\phi-\phi_J$.
The critical exponents are given by $\nu=1.0$ and $\lambda=4.0$, where slopes of the dotted and solid lines are $-\nu/\lambda$ and $-0.3$, respectively.
\label{fig:diff}}
\end{figure}

To quantify the normal diffusion of the disks, we introduce the diffusivity (or diffusion coefficient) as
\footnote{We define the diffusivity [Eq.\ (\ref{eq:D})] as the slope of the MSD [Eq.\ (\ref{eq:MSD})] in the normal diffusive regime $\gamma=\dot{\gamma}\tau>1$,
where we take sample averages of $\Delta(\tau)^2/2\tau$ as $D/\dot{\gamma}\equiv <\Delta(\gamma)^2/2\gamma>$ in the range between $1<\gamma<10^2$.}
\begin{equation}
D=\lim_{\tau\rightarrow\infty}\frac{\Delta(\tau)^2}{2\tau}~.
\label{eq:D}
\end{equation}
Figure \ref{fig:diff}(a) shows double logarithmic plots of the diffusivity [Eq.\ (\ref{eq:D})] over the shear rate $D/\dot{\gamma}$,
where symbols represent the packing fraction $\phi$ (as listed in the legend).
The diffusivity over the shear rate increases with $\phi$.
If the system is above jamming $\phi>\phi_J$, it is a monotonously decreasing function of $\dot{\gamma}$.
On the other hand, if the system is below jamming $\phi<\phi_J$, it exhibits a crossover from plateau to a monotonous decrease
around a characteristic shear rate,\ e.g.\ $\dot{\gamma}_0t_0\simeq 10^{-3}$ for $\phi=0.80$ \cite{diff_shear_md3,diff_shear_md4}.

In Appendix \ref{sec:rheo}, we have demonstrated \emph{scaling collapses} of rheological flow curves \cite{rheol0}.
Here, we also demonstrate scaling collapses of the diffusivity.
As shown in Fig.\ \ref{fig:diff}(b), all the data are nicely collapsed
\footnote{The data for the highest shear rate, $\dot{\gamma}=10^{-1}t_0^{-1}$, is removed from the scaling collapses in Figs.\ \ref{fig:diff}(b) and \ref{fig:xi}(a).}
by the scaling exponents, $\nu=1.0$ and $\lambda=4.0$.
If the shear rate is smaller than a characteristic value as $\dot{\gamma}/|\Delta\phi|^\lambda \lesssim 10^4$,\
i.e.\ $\dot{\gamma}<\dot{\gamma}_c\simeq 10^4|\Delta\phi|^\lambda$, the data below jamming ($\phi<\phi_J$) are constant.
However, the data above jamming ($\phi>\phi_J$) show the power-law decay (solid line), where the slope is $-0.305 \pm 0.002$ (approximately $-0.3$).
Therefore, we describe the diffusivity in a \emph{quasi-static regime} ($\dot{\gamma}<\dot{\gamma}_c$)
as $|\Delta\phi|^\nu D/\dot{\gamma}\sim\mathcal{G}_\pm(\dot{\gamma}/|\Delta\phi|^\lambda)$,
where the scaling functions are given by $\mathcal{G}_-(x)\sim\mathrm{const.}$ for $\phi<\phi_J$ and $\mathcal{G}_+(x)\sim x^{-0.3}$ otherwise.
On the other hand, if $\dot{\gamma}>\dot{\gamma}_c$, all the data follow a single power law (dotted line).
This means that the scaling functions are given by $\mathcal{G}_\pm(x) \sim x^{-z}$ in a \emph{plastic flow regime} ($\dot{\gamma}>\dot{\gamma}_c$),
where the diffusivity scales as $D\sim\dot{\gamma}|\Delta\phi|^{-\nu}\mathcal{G}_\pm(\dot{\gamma}/|\Delta\phi|^\lambda)\sim\dot{\gamma}^{1-z}|\Delta\phi|^{\lambda z-\nu}$.
Because this scaling should be independent of whether the system is below or above jamming,\
i.e.\ independent of $|\Delta\phi|$, the power-law exponent is given by $z=\nu/\lambda=1/4$ as confirmed in Fig.\ \ref{fig:diff}(b).
%

In summary, the diffusivity of the disks scales as
\begin{equation}
D \sim
\begin{cases}
|\Delta\phi|^{-\nu}\dot{\gamma} & (\phi<\phi_J) \\
|\Delta\phi|^{0.3\lambda-\nu}\dot{\gamma}^{0.7} & (\phi>\phi_J)
\end{cases} \label{eq:D1}
\end{equation}
in the quasi-static regime ($\dot{\gamma}<\dot{\gamma}_c$) and
\begin{equation}
D \sim \dot{\gamma}^{1-\nu/\lambda}
\label{eq:D2}
\end{equation}
in the plastic flow regime ($\dot{\gamma}>\dot{\gamma}_c$), where the critical exponents are estimated as $\nu=1.0$ and $\lambda=4.0$.
From Eqs.\ (\ref{eq:D1}) and (\ref{eq:D2}), we find that the diffusivity below jamming ($\phi<\phi_J$) is linear in the shear rate $D\sim\dot{\gamma}$ in slow flows,
whereas its dependence on the shear rate is algebraic $D\sim\dot{\gamma}^{3/4}$ in fast flows.
A similar trend has been found in molecular dynamics studies of simple shear flows below jamming \cite{diff_shear_md0,dh_md2,dh_md1}.
In addition, the proportionality for the diffusivity below jamming diverges at the transition as $|\Delta\phi|^{-1}$ [Eq.\ (\ref{eq:D1})],
which we will relate to a length scale diverging as the system approaches jamming from below (Sec.\ \ref{sub:rigid}).
The diffusivity above jamming ($\phi>\phi_J$) implies the crossover from $D\sim|\Delta\phi|^{0.2}\dot{\gamma}^{0.7}$ to $\dot{\gamma}^{3/4}=\dot{\gamma}^{0.75}$,
which reasonably agrees with the prior work on soft athermal disks under shear \cite{diff_shear_md0}.
Interestingly, the crossover shear rate vanishes at the transition as $\dot{\gamma}_c\sim|\Delta\phi|^{4.0}$,
which is reminiscent of the fact that the crossover from the Newtonian or yield stress to the plastic flow vanishes at the onset of jamming (see Appendix \ref{sec:rheo}).
%
\subsection{Rigid clusters}
\label{sub:rigid}
\begin{figure}
\includegraphics[width=\columnwidth]{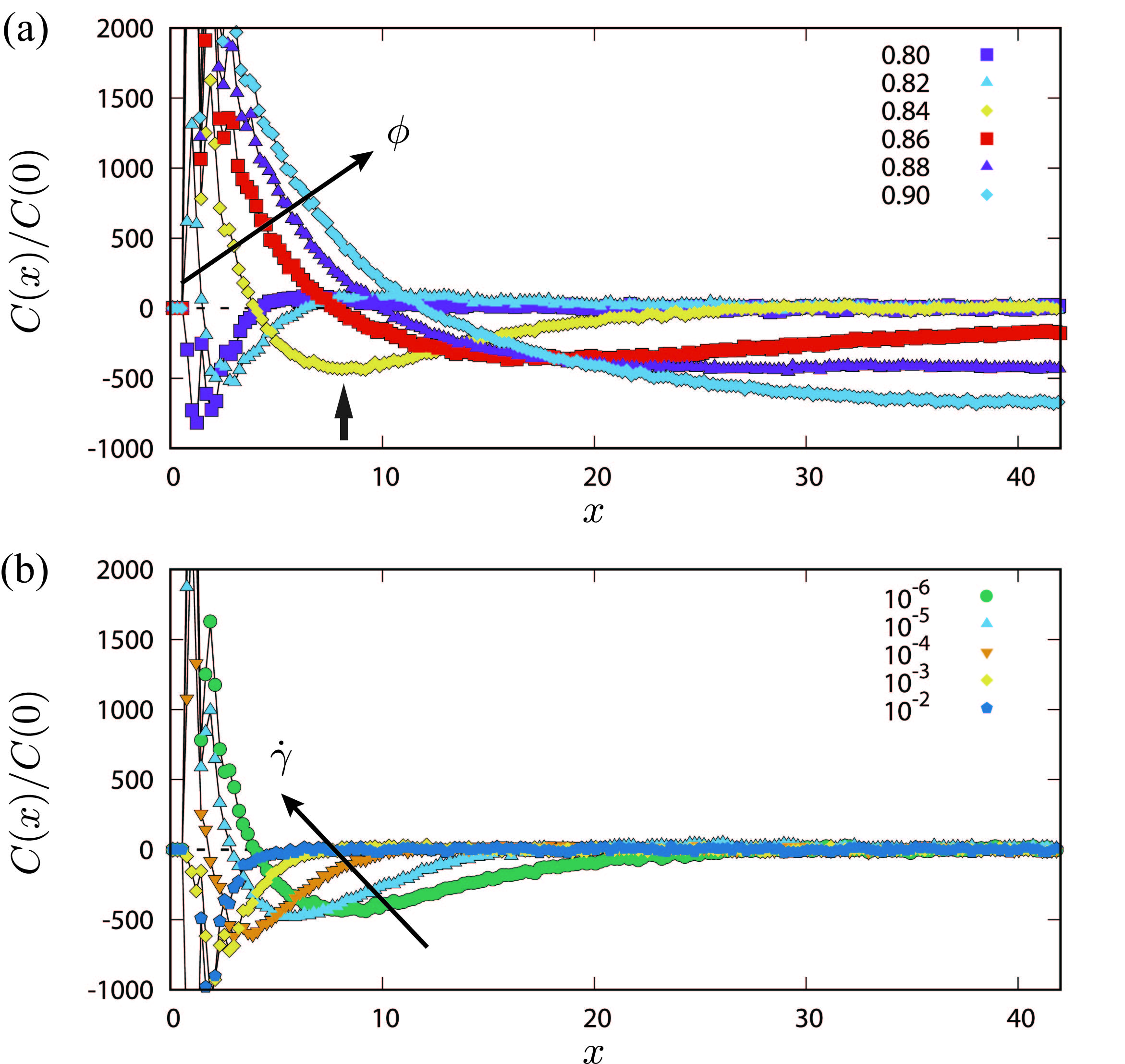}
\caption{
Normalized spatial correlation functions of the transverse velocities $C(x)/C(0)$, where symbols are as in Fig.\ \ref{fig:msdy}.
(a) The packing fraction $\phi$ increases as indicated by the arrow and listed in the legend, where $\dot{\gamma}=10^{-6}t_0^{-1}$.
The minimum of the data for $\phi=0.84$ is indicated by the vertical (gray) arrow.
(b) The shear rate $\dot{\gamma}$ increases as indicated by the arrow and listed in the legend, where $\phi=0.84$.
\label{fig:corl}}
\end{figure}

We now relate the diffusivity $D$ to rigid clusters of soft athermal particles under shear (Fig.\ \ref{fig:rigid}).
The rigid clusters represent collective motions of the particles which tend to move in the same direction \cite{saitoh11}.
According to the literature of jamming \cite{rheol0,pdf1,corl3}, we quantify the collective motions by a spatial correlation function
$C(x)=\langle v_y(x_i,y_i)v_y(x_i+x,y_i)\rangle$, where $v_y(x,y)$ is the transverse velocity field
and the ensemble average $\langle\dots\rangle$ is taken over disk positions and time (in a steady state).
Figure \ref{fig:corl} shows the normalized correlation function $C(x)/C(0)$, where the horizontal axis ($x$-axis) is scaled by the mean disk diameter $d_0$.
As can be seen, the correlation function exhibits a well-defined minimum at a characteristic length scale $x=\xi$
(as indicated by the vertical arrow for the case of $\phi=0.84$ in Fig.\ \ref{fig:corl}(a)).
Because the minimum is negative $C(\xi)<0$, the transverse velocities are most ``anti-correlated" at $x=\xi$ \cite{rheol0}.
Therefore, if we assume that the rigid clusters are circular, their mean diameter is comparable in size with $\xi$
\footnote{
In a short range of $x$, non-affine velocities tend to be aligned in the same direction,\ i.e.\ $v_y(x_i,y_i)v_y(x_i+x,y_i)>0$,
such that the correlation function $C(x)$ is a positive decreasing function of $x$.
If the non-affine velocities start to align in the opposite direction,\ i.e.\ $v_y(x_i,y_i)v_y(x_i+x,y_i)<0$, the correlation function becomes negative, $C(x)<0$.
Because the correlation function is minimum when the non-affine velocities are located on either side of a vortex-like structure,
a typical size of rigid clusters can be defined as the distance at which $C(x)$ becomes minimum.
In a long distance of $x$, due to the randomness of non-affine velocities, the correlation function eventually decays to zero with or without oscillations \cite{saitoh11}.}.
The length scale $\xi$ increases with the increase of $\phi$ [Fig.\ \ref{fig:corl}(a)] but decreases with the increase of $\dot{\gamma}$ [Fig.\ \ref{fig:corl}(b)].
These results are consistent with the fact that the collective behavior is most enhanced in slow flows of dense systems \cite{saitoh11}.
%

\begin{figure}
\includegraphics[width=\columnwidth]{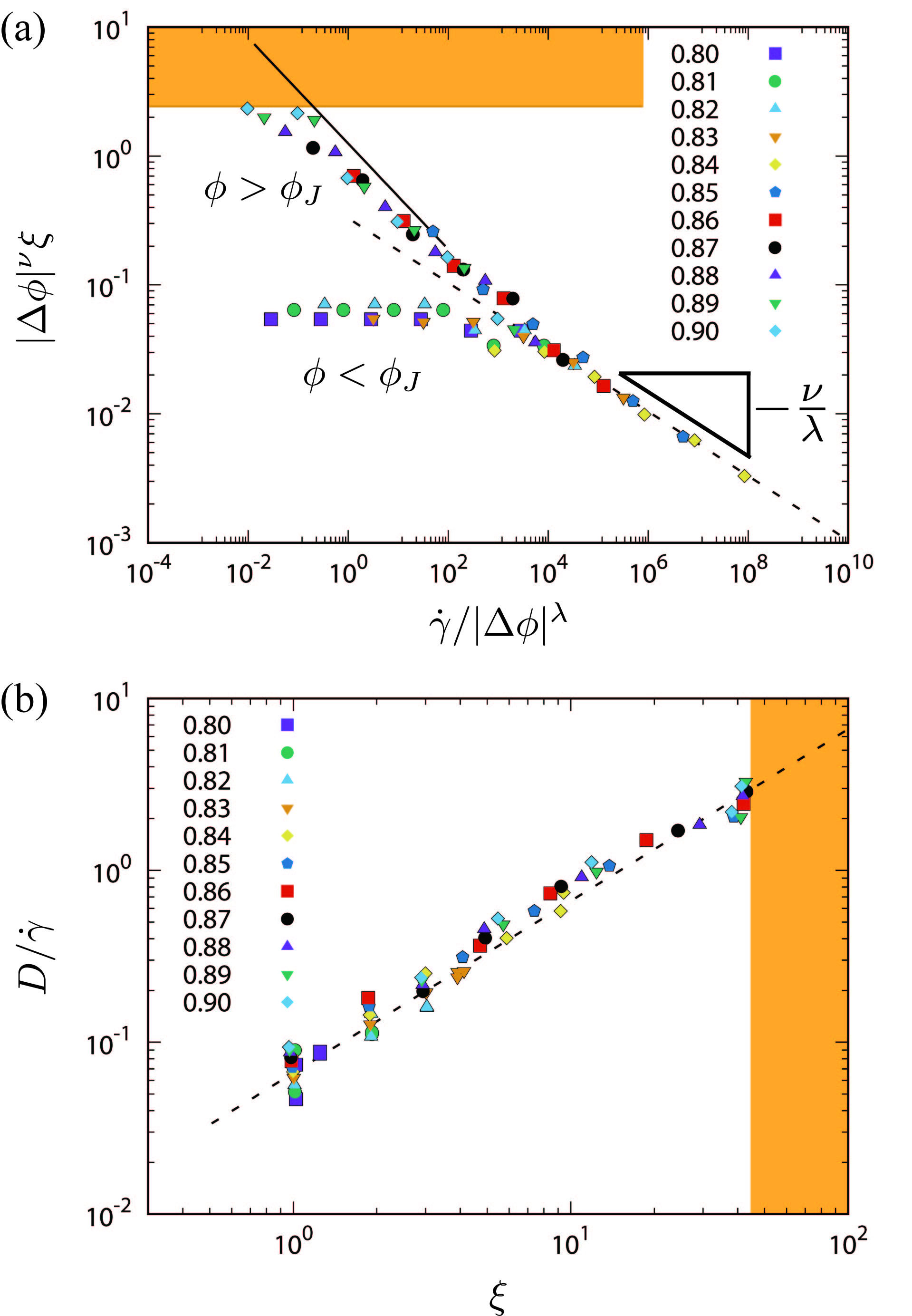}
\caption{
(a) Scaling collapses of the length scale $\xi$, where $\Delta\phi\equiv\phi-\phi_J$ and $\phi$ increases as listed in the legend.
The critical exponents are $\nu=1.0$ and $\lambda=4.0$ as in Fig.\ \ref{fig:diff}(b),
where slopes of the dotted and solid lines are given by $-\nu/\lambda$ and $-0.4$, respectively.
The shaded region exceeds the system size $|\Delta\phi|^\nu L/2$ for the case of $\phi=0.90$.
(b) Scatter plots of the diffusivity over the shear rate $D/\dot{\gamma}$ and the length scale $\xi$, where $\phi$ increases as listed in the legend.
The dotted line represents a linear relation $D/\dot{\gamma}\sim\xi$ and the shaded region exceeds the system size $L/2\simeq 44d_0$.
\label{fig:xi}}
\end{figure}

As reported in Ref.\ \cite{rheol0}, we examine critical scaling of the length scale.
Figure \ref{fig:xi}(a) displays scaling collapses of the data of $\xi$, where the critical exponents, $\nu=1.0$ and $\lambda=4.0$, are the same with those in Fig.\ \ref{fig:diff}(b).
If the shear rate is smaller than the characteristic value,\ i.e.\ $\dot{\gamma}<\dot{\gamma}_c\simeq 10^4|\Delta\phi|^\lambda$,
the data below jamming ($\phi<\phi_J$) exhibit plateau, whereas those above jamming ($\phi>\phi_J$) diverge with the \emph{decrease} of shear rate.
Therefore, if we assume that the data above jamming follow the power-law (solid line) with the slope $-0.40 \pm 0.02$ (approximately $-0.4$),
the length scale in the quasi-static regime ($\dot{\gamma}<\dot{\gamma}_c$) can be described as
$|\Delta\phi|^\nu\xi\sim\mathcal{J}_\pm(\dot{\gamma}/|\Delta\phi|^\lambda)$ with the scaling functions,
$\mathcal{J}_-(x)\sim\mathrm{const.}$ for $\phi<\phi_J$ and $\mathcal{J}_+(x)\sim x^{-0.4}$ otherwise.
Note that, however, the length scale is limited to the system size $L$ [shaded region in Fig.\ \ref{fig:xi}(a)]
and should be scaled as $\xi\sim L$ above jamming in the quasi-static limit $\dot{\gamma}\rightarrow 0$ \cite{pdf1,diff_shear_md3,nafsc2}.
This means that the system size is the only relevant length scale \cite{nafsc0} and thus we conclude $\xi\sim L$ in slow flows of jammed systems.
On the other hand, if $\dot{\gamma}>\dot{\gamma}_c$, all the data are collapsed onto a single power law [dotted line in Fig.\ \ref{fig:xi}(a)].
Therefore, the scaling functions are given by $\mathcal{J}_\pm(x)\sim x^{-z}$ such that the length scale scales as $\xi\sim\dot{\gamma}^{-z}|\Delta\phi|^{\lambda z-\nu}$.
Because this relation is independent of $|\Delta\phi|$, the exponent should be $z=\nu/\lambda$ as confirmed in Fig.\ \ref{fig:xi}(a).
%

In summary, the length scale, or the mean size of rigid clusters, scales as
\begin{equation}
\xi \sim
\begin{cases}
|\Delta\phi|^{-\nu} & (\phi<\phi_J) \\
L & (\phi>\phi_J)
\end{cases} \label{eq:xi1}
\end{equation}
in the quasi-static regime ($\dot{\gamma}<\dot{\gamma}_c$) and
\begin{equation}
\xi \sim \dot{\gamma}^{-\nu/\lambda}
\label{eq:xi2}
\end{equation}
in the plastic flow regime ($\dot{\gamma}>\dot{\gamma}_c$),
where the critical exponents, $\nu$ and $\lambda$, are the same with those for the diffusivity [Eqs.\ (\ref{eq:D1}) and (\ref{eq:D2})].
The critical divergence below jamming in the quasi-static regime,\ i.e.\ $\xi\sim|\Delta\phi|^{-1}$ [Eq.\ (\ref{eq:xi1})],
is consistent with the results of quasi-static simulations ($\dot{\gamma}\rightarrow 0$) of sheared athermal disks \cite{nafsc2,dh_qs1}.
In addition, the scaling $\xi\sim\dot{\gamma}^{-1/4}$ in the plastic flow regime [Eq.\ (\ref{eq:xi2})] is very close to the prior work on athermal particles under shear \cite{dh_md2}.

From the results of the diffusivity [Eqs.\ (\ref{eq:D1}) and (\ref{eq:D2})] and length scale [Eqs.\ (\ref{eq:xi1}) and (\ref{eq:xi2})],
we discuss how the rigid clusters contribute to the diffusion of the particles.
The linear relation $D\sim d_0\xi\dot{\gamma}$ [Eq.\ (\ref{eq:rigid_cluster})] holds below jamming (regardless of $\dot{\gamma}$) and in the plastic flow regime (regardless of $\phi$).
We stress that the divergence of the diffusivity over the shear rate in the quasi-static regime,\ i.e.\ $D/\dot{\gamma}\sim|\Delta\phi|^{-1}$ [Eq.\ (\ref{eq:D1})],
is caused by the diverging length scale below jamming,\ i.e.\ $\xi\sim|\Delta\phi|^{-1}$ [Eq.\ (\ref{eq:xi1})].
As shown in Fig.\ \ref{fig:xi}(b), the linear relation (dotted line) well explains our results if the length scale $\xi$ is smaller than $10d_0$.
If the system is above jamming, the length scale increases (more than $10d_0$) with the increase of $\phi$.
However, the diffusivity over the shear rate $D/\dot{\gamma}$ starts to deviate from the linear relation (dotted line) and the length scale reaches the system size $L/2\simeq 44d_0$ (shaded region).
We conclude that this deviation is caused by finite-size effects and further studies of different system sizes are necessary (as in Refs.\ \cite{diff_shear_md3,diff_shear_md4})
to figure out the relation between $D/\dot{\gamma}$ and $\xi$ in this regime, which we postpone as a future work.
%
\section{Discussions}
\label{sec:disc}
In this study, we have numerically investigated rheological and transport properties of soft athermal particles under shear.
Employing MD simulations of two-dimensional disks, we have clarified how the rheology, self-diffusion, and size of rigid clusters vary with the control parameters,\
i.e.\ the externally imposed shear rate $\dot{\gamma}$ and packing fraction of the disks $\phi$.

Our main result is the critical scaling of the diffusivity (Sec.\ \ref{sub:diff}),
where their dependence on both $\dot{\gamma}$ and $\phi$ is reported [Eqs.\ (\ref{eq:D1})-(\ref{eq:xi2})].
The diffusivity has been calculated on both sides of jamming (by a single numerical protocol) to unify the understanding of self-diffusion of soft particulate systems.
We found that (i) the diffusivity below jamming exhibits a crossover from the linear scaling $D\sim\dot{\gamma}$ to the power-law $D\sim\dot{\gamma}^{3/4}$.
Such a crossover can also be seen in previous simulations \cite{diff_shear_md7,diff_shear_md6,dh_md2}.
In addition, (ii) the diffusivity below jamming diverges as $D\sim|\Delta\phi|^{-1}$ if the system is in the quasi-static regime ($\dot{\gamma}<\dot{\gamma}_c$),
whereas (iii) the diffusivity (both below and above jamming) is insensitive to $\phi$ if the system is in the plastic flow regime ($\dot{\gamma}>\dot{\gamma}_c$).
Note that (iv) the crossover shear rate vanishes at the onset of jamming as $\dot{\gamma}_c\sim|\Delta\phi|^{4.0}$.
These results (ii)-(iv) are the new findings of this study.
On the other hand, we found that (v) the diffusivity above jamming is weakly dependent on $\phi$ (as $D\sim|\Delta\phi|^{0.2}$) in the quasi-static regime
and (vi) shows a slight change from $D\sim\dot{\gamma}^{0.7}$ to $\dot{\gamma}^{3/4}$.
Though the result (v) is the new finding, the result (vi) contrasts with the prior studies,
where the diffusivity exhibits a crossover from $D\sim\dot{\gamma}$ to $\dot{\gamma}^{1/2}$ \cite{diff_shear_md1,diff_shear_md3,diff_shear_md4}.
Because our scaling $D\sim\dot{\gamma}^{0.7}$ in the quasi-static regime is consistent with Ref.\ \cite{diff_shear_md0}, where the same overdamped dynamics are used,
we suppose that the discrepancy is caused by numerical models and flow conditions,
where the Lennard-Jones potential is used for the interaction between the particles in Refs.\ \cite{diff_shear_md3,diff_shear_md4}
and frictionless/frictional disks are sheared under constant pressure in Ref.\ \cite{diff_shear_md1}.

We have also examined the relation between the diffusivity and typical size of rigid clusters $\xi$ (Sec.\ \ref{sub:rigid}).
Below jamming, we found the critical divergence $\xi\sim|\Delta\phi|^{-1}$ in the quasi-static regime
as previously observed in quasi-static simulations ($\dot{\gamma}\rightarrow 0$) of sheared athermal disks \cite{nafsc2}.
Note that our exponent is larger than that obtained from previous studies of soft athermal particles \cite{lng0,rheol0},
where the critical divergence $\xi\sim|\Delta\phi|^{-\nu}$ is characterized by the exponent in the range between $0.6<\nu<0.7$.
Though the meaning of $\xi$ is similar, in the sense that it quantifies the size of collective motions of the particles under shear,
we found that the quality of data collapse of $\xi$ is even worse if we use the exponents $\nu=0.6$ and $0.7$.
So far, we cannot figure out the cause of this discrepancy and postpone this problem as a future work.
In the plastic flow regime, the size becomes independent of $\phi$ and scales as $\xi\sim\dot{\gamma}^{-1/4}$.
This is consistent with the previous result of sheared athermal particles \cite{dh_md2} (and is also close to the result of thermal glasses under shear \cite{th-dh_md1}).
Above jamming, however, the size exhibits a crossover from $\xi\sim L$ to $\dot{\gamma}^{-1/4}$
which contrasts with the crossover from $\xi\sim\mathrm{const.}$ to $\dot{\gamma}^{-1/2}$ previously reported in simulations of amorphous solids \cite{diff_shear_md1,diff_shear_md3,diff_shear_md4}.
From our scaling analyses, we found that the linear relation $D\sim d_0\xi\dot{\gamma}$ [Eq.\ (\ref{eq:rigid_cluster})] holds
below jamming (for $\forall\dot{\gamma}$) and in the plastic flow regime (for $\forall\phi$),
indicating that the self-diffusion is enhanced by the rotation of rigid clusters \cite{rheol0,diff_shear_md1}.

In our MD simulations, we fixed the system size to $L\simeq 88d_0$.
However, systematic studies of different system sizes are needed to clarify the relation between $D$ and $\xi\sim L$ above jamming,
especially in the quasi-static limit $\dot{\gamma}\rightarrow 0$ \cite{diff_shear_md3,diff_shear_md4}.
In addition, our analyses are limited to two dimensions.
Though previous studies suggest that the diffusivity is independent of the dimensionality \cite{diff_shear_md7,diff_shear_md6,dh_md2},
a recent study of soft athermal particles reported that the critical scaling of shear viscosity depends on dimensions \cite{rheol15}.
Therefore, it is important to check whether the critical scaling [Eqs.\ (\ref{eq:D1}) and (\ref{eq:D2})] is different in three-dimensional systems.
In addition, O'Hern et al. found that the critical exponents near jamming are sensitive to interaction potentials \cite{gn1}.
Therefore, it is important to examine whether our exponents, $\nu$ and $\lambda$, are sensitive to the details of numerical models,\
e.g.\ contact forces, particle inertia, microscopic friction, particle size distributions,\ etc.\ with the aim of testing \emph{universality class} of the diffusivity and size of rigid clusters.
Moreover, the analysis of local structures of instantaneous configurations of the disks is important to investigate the rigidity of the clusters \cite{q6shear} and
the relation between the diffusivity and shear viscosity may be interesting because it gives a Stokes-Einstein like relation for the non-equilibrium systems studied here.

In our recent study of non-Brownian suspensions \cite{rheol11}, we found that the viscosity exhibits a crossover when the system approaches the jamming transition.
To describe such a crossover, a correction term should be added to the conventional scaling function as $\mathcal{F}(x)+\mathcal{G}(x)$,
where two different exponents are introduced as $\mathcal{F}(x)\sim x^{-z}$ and $\mathcal{G}(x)\sim x^{-w}$.
This means that the scaling collapse of the flow curves \cite{rheol0}, where the scaling function is given by a single power law, can be violated if the system is very close to jamming.
Similarly, our scaling collapses of $D$ [Fig.\ \ref{fig:diff}(b)] and $\xi$ [Fig.\ \ref{fig:xi}(a)] could break down by the same reason.
Therefore, it is necessary to examine whether the crossovers of $D$ and $\xi$ exist in the case that the system is in the vicinity of the jamming transition.
%
\begin{acknowledgments}
We thank B. P. Tighe, L. Berthier, H. Hayakawa, M. Otsuki, and S. Takada for fruitful discussions.
K.S. thanks F. Radjai and W. Kob for fruitful discussions and warm hospitality in Montpellier.
This work was supported by KAKENHI Grant No.\ 16H04025, No.\ 18K13464 and No.\ 19K03767 from JSPS.
\end{acknowledgments}
%
\appendix
\section{Rheology}
\label{sec:rheo}
\begin{figure}
\includegraphics[width=\columnwidth]{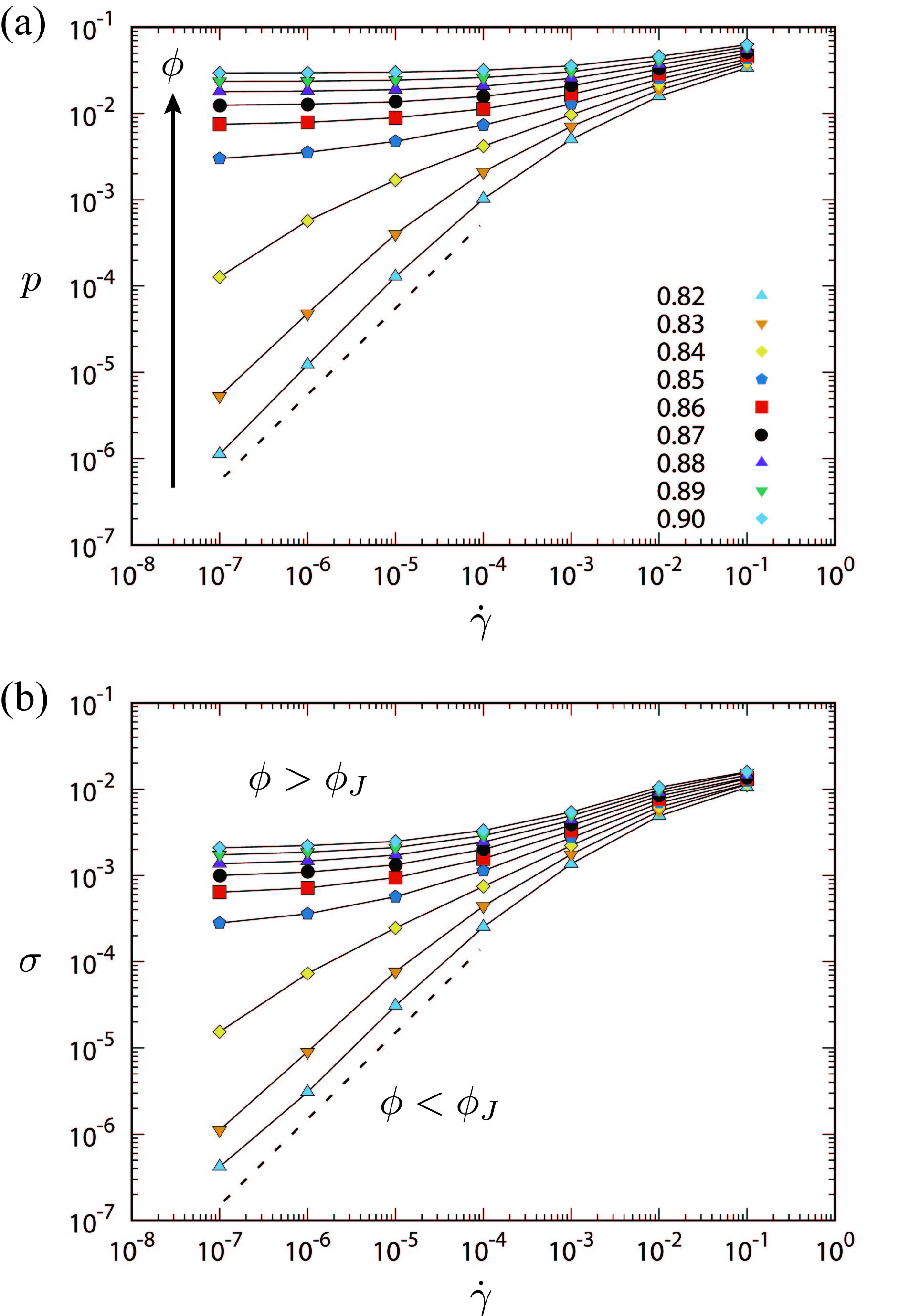}
\caption{
\emph{Flow curves},\ i.e.\ (a) the pressure $p$ and (b) shear stress $\sigma$ as functions of the shear rate $\dot{\gamma}$.
The area fraction $\phi$ increases as indicated by the arrow (listed in the legend) in (a).
The dotted lines represent the Newtonian behavior,\ i.e.\ (a) $p\sim\dot{\gamma}$ and (b) $\sigma\sim\dot{\gamma}$,
for low area fractions, $\phi<\phi_J$, where $\phi_J\simeq 0.8433$ is the jamming transition density.
\label{fig:rheo}}
\end{figure}

The rheology of soft athermal particles is dependent on both the shear rate $\dot{\gamma}$ and area fraction $\phi$ \cite{rheol0,pdf1,rheol7}.
Figure \ref{fig:rheo} displays our numerical results of \emph{flow curves},\ i.e.\ (a) the pressure $p$ and (b) shear stress $\sigma$ as functions of the shear rate $\dot{\gamma}$.
Here, different symbols represent different values of $\phi$ (as listed in the legend of (a)).
The pressure and shear stress are defined as $p=(\tau_{xx}+\tau_{yy})/2$ and $\sigma=-\tau_{xy}$, respectively, where the stress tensor is given by the virial expression
\begin{equation}
\tau_{\alpha\beta}=\frac{1}{L^2}\sum_i\sum_{j~(>i)}f_{ij\alpha}^\mathrm{e}r_{ij\beta}
\label{eq:stress}
\end{equation}
($\alpha,\beta=x,y$)
with the $\alpha$-component of elastic force $f_{ij\alpha}^\mathrm{e}$ and the $\beta$-component of relative position $r_{ij\beta}$.
As shown in Fig.\ \ref{fig:rheo}, both the pressure and shear stress exhibit the Newtonian behavior,\
i.e.\ they are proportional to the shear rate, $p\sim\dot{\gamma}$ and $\sigma\sim\dot{\gamma}$ (dotted lines),
only if the area fraction is lower than the jamming transition density ($\phi<\phi_J$) and the shear rate is small enough ($\dot{\gamma}t_0\lesssim 10^{-4}$).
However, a finite yield stress, $p_Y>0$ and $\sigma_Y>0$, emerges in the zero shear limit $\dot{\gamma}\rightarrow 0$ if the system is above jamming ($\phi>\phi_J$).
%

\begin{figure}
\includegraphics[width=\columnwidth]{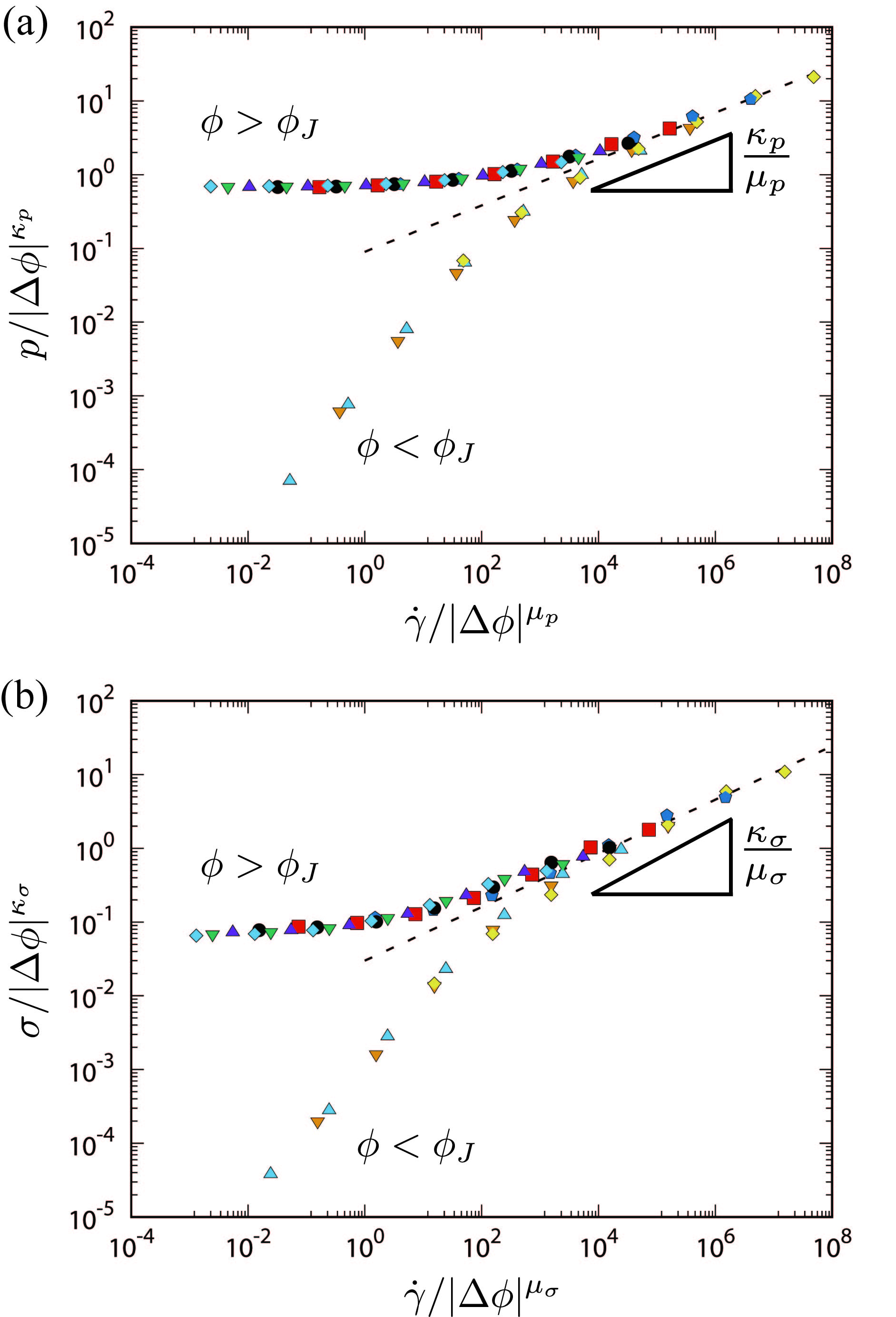}
\caption{
Scaling collapses of (a) the pressure and (b) shear stress, where $\Delta\phi\equiv\phi-\phi_J$ is the proximity to jamming.
See the text for critical exponents, $\kappa_p$, $\mu_p$, $\kappa_\sigma$, and $\mu_\sigma$, where the dotted lines have the slopes (a) $\kappa_p/\mu_p$ and (b) $\kappa_\sigma/\mu_\sigma$.
\label{fig:rheo-clp}}
\end{figure}

In the literature of jamming \cite{rheol0,pdf1,rheol7}, rheological flow curves are collapsed by critical scaling.
This means that the crossover from the Newtonian behavior ($p\sim\dot{\gamma}$ and $\sigma\sim\dot{\gamma}$) or the yield stress ($p\sim p_Y$ and $\sigma\sim\sigma_Y$)
to plastic flow regime vanishes as the system approaches jamming $\phi\rightarrow\phi_J$.
To confirm this trend, we collapse the data in Fig.\ \ref{fig:rheo} by the proximity to jamming $|\Delta\phi|\equiv|\phi-\phi_J|$ as in Fig.\ \ref{fig:rheo-clp}.
Though the critical exponents are slightly different,\ i.e.\ $\kappa_p=1.1$ and $\mu_p=3.5$ for the pressure [Fig.\ \ref{fig:rheo-clp}(a)]
and $\kappa_\sigma=1.2$ and $\mu_\sigma=3.3$ for the shear stress [Fig.\ \ref{fig:rheo-clp}(b)], all the data are nicely collapsed on top of each other.
If the shear rate is small enough, the data below jamming ($\phi<\phi_J$) follow the lower branch, whereas the data above jamming ($\phi>\phi_J$) are almost constant.
Therefore, the pressure and shear stress can be described as $p/|\Delta\phi|^{\kappa_p}\sim\mathcal{F}_\pm(\dot{\gamma}/|\Delta\phi|^{\mu_p})$
and $\sigma/|\Delta\phi|^{\kappa_\sigma}\sim\mathcal{F}_\pm(\dot{\gamma}/|\Delta\phi|^{\mu_\sigma})$ with the scaling functions,
$\mathcal{F}_-(x)\sim x$ for $\phi<\phi_J$ and $\mathcal{F}_+(x)\sim\mathrm{const.}$ for $\phi>\phi_J$.

On the other hand, if the shear rate is large enough, the system is in plastic flow regime,
where all the data (both below and above jamming) follow a single power law (dotted lines in Fig.\ \ref{fig:rheo-clp}).
This implies that the scaling functions are given by $\mathcal{F}_\pm(x)\sim x^z$ (for both $\phi<\phi_J$ and $\phi>\phi_J$) with a power-law exponent $z$.
Then, the pressure and shear stress scale as $p\sim|\Delta\phi|^{\kappa_p}\mathcal{F}_\pm(\dot{\gamma}/|\Delta\phi|^{\mu_p})\sim\dot{\gamma}^z|\Delta\phi|^{\kappa_p-\mu_p z}$
and $\sigma\sim|\Delta\phi|^{\kappa_\sigma}\mathcal{F}_\pm(\dot{\gamma}/|\Delta\phi|^{\mu_\sigma})\sim\dot{\gamma}^z|\Delta\phi|^{\kappa_\sigma-\mu_\sigma z}$, respectively.
These scaling relations should be independent of whether the system is below or above jamming,\ i.e.\ independent of $|\Delta\phi|$.
Thus, the power-law exponent is $z=\kappa_p/\mu_p\simeq 0.31$ for the pressure and $z=\kappa_\sigma/\mu_\sigma\simeq 0.36$ for the shear stress as confirmed in Fig.\ \ref{fig:rheo-clp} (dotted lines).
Note that the scaling collapses in Fig.\ \ref{fig:rheo-clp} also confirm that the jamming transition density $\phi_J\simeq 0.8433$ is correct in our sheared systems \cite{rheol0}.
%

In summary, the rheological flow properties of the disks are described as
\begin{eqnarray}
p &\sim&
\begin{cases}
|\Delta\phi|^{\kappa_p-\mu_p}\dot{\gamma} & (\phi<\phi_J) \\
|\Delta\phi|^{\kappa_p} & (\phi>\phi_J)
\end{cases}~, \label{eq:pressure1} \\
\sigma &\sim&
\begin{cases}
|\Delta\phi|^{\kappa_\sigma-\mu_\sigma}\dot{\gamma} & (\phi<\phi_J) \\
|\Delta\phi|^{\kappa_\sigma} & (\phi>\phi_J)
\end{cases}~, \label{eq:shear_stress1}
\end{eqnarray}
in the quasi-static regime and
\begin{eqnarray}
p &\sim& \dot{\gamma}^{\kappa_p/\mu_p}~, \label{eq:pressure2} \\
\sigma &\sim& \dot{\gamma}^{\kappa_\sigma/\mu_\sigma}~, \label{eq:shear_stress2}
\end{eqnarray}
in the plastic flow regime.
The critical exponents are estimated as $\kappa_p=1.1$, $\mu_p=3.5$, $\kappa_\sigma=1.2$, and $\mu_\sigma=3.3$.
In Eqs.\ (\ref{eq:pressure1}) and (\ref{eq:shear_stress1}), the Newtonian behavior is given by $p\sim|\Delta\phi|^{-2.4}\dot{\gamma}$ and $\sigma\sim|\Delta\phi|^{-2.1}\dot{\gamma}$,
where the exponents are comparable to those for viscosity divergence below jamming \cite{rheol7}.
The yield stress vanishes as $p_Y\sim|\Delta\phi|^{1.1}$ and $\sigma_Y\sim|\Delta\phi|^{1.2}$ when the system approaches jamming from above [Eqs.\ (\ref{eq:pressure1}) and (\ref{eq:shear_stress1})],
which is consistent with the previous study of two-dimensional bubbles under shear \cite{pdf1}.
The scaling in the plastic flow regime, $p\sim\dot{\gamma}^{0.31}$ and $\sigma\sim\dot{\gamma}^{0.36}$ [Eqs.\ (\ref{eq:pressure2}) and (\ref{eq:shear_stress2})],
is close to the prior work on sheared athermal disks \cite{rheol14}, indicating \emph{shear thinning} as typical for particulate systems under shear \cite{larson}.
%
\section{Non-affine displacements}
\label{sec:nona}
The self-diffusion of soft athermal particles is also sensitive to both $\dot{\gamma}$ and $\phi$.
Because our system is homogeneously sheared (along the $x$-direction), the self-diffusion is represented by fluctuating motions of the disks around a mean flow.
In our MD simulations, the mean velocity field is determined by the affine deformation as $\dot{\gamma}y\bm{e}_x$, where $\bm{e}_x$ is a unit vector parallel to the $x$-axis.
Therefore, subtracting the mean velocity field from each disk velocity $\bm{u}_i(t)$, we introduce non-affine velocities as $\Delta\bm{u}_i(t)=\bm{u}_i(t)-\dot{\gamma}y_i\bm{e}_x$ ($i=1,\dots,N$).
\emph{Non-affine displacements} are then defined as the time integrals
\begin{equation}
\Delta\bm{r}_i(\tau) = \int_{t_a}^{t_a+\tau}\Delta\bm{u}_i(t)dt~,
\label{eq:non-affine}
\end{equation}
where $\tau$ is the time interval.
Note that the initial time $t_a$ can be arbitrary chosen during a steady state.

It is known that the non-affine displacements [Eq.\ (\ref{eq:non-affine})] are sensitive to the rheological flow properties (Sec.\ \ref{sec:rheo}) \cite{saitoh11}.
Their magnitude significantly increases if the packing fraction exceeds the jamming point.
In addition, their spatial distributions become more ``collective" (they tend to align in the same directions with neighbors) with the decrease of the shear rate.
This means that the self-diffusion is also strongly dependent on both the shear rate and density.
Especially, the collective behavior of the non-affine displacements implies the growth of rigid clusters in slow flows $\dot{\gamma}t_0\ll 1$ of jammed systems $\phi>\phi_J$,
where the yield stress $\sigma\sim\sigma_Y$ is observed in the flow curves (Fig.\ \ref{fig:rheo}).
%
\bibliography{diffusion_overdamp}
\end{document}